\documentstyle[multicol,prl,aps,graphics]{revtex}
\begin{document}
\title{Violation of Ioffe-Regel condition but saturation of 
resistivity of the high $T_c$ cuprates}
\author{M. Calandra and O. Gunnarsson}
\address{ Max-Planck-Institut f\"ur Festk\"orperforschung 
D-70506 Stuttgart, Germany}

\maketitle

\begin{abstract}
We demonstrate that the resistivity data of a number of high $T_c$ cuprates,
in particular La$_{2-x}$Sr$_x$CuO$_4$, are consistent with resistivity 
saturation, although the Ioffe-Regel condition is strongly violated. 
By using the f-sum rule together with calculations
of the kinetic energy in the $t-J$ model, we show that the saturation
resistivity is unusually large. This is related to the strong reduction
of the kinetic energy due to strong correlation effects. The fulfilment
of the Ioffe-Regel condition for conventional transition metal compounds 
is found to be somewhat accidental.
\end{abstract}

\begin{multicols}{2}
Some strongly correlated metals\cite{laruo,sr2ruo4,cro,srruo3}, in particular
many high $T_c$ 
cuprates\cite{Gurvitch,hightcres,Ando,ybacuo,ndcecuo,bisrcaycuo,steglich,tlbacuo,hgbacacuo,Allenhightc},
show an exceptionally large resistivity, $\rho$, at high temperatures $T$.
Thus $\rho$ can reach values of several m$\Omega$cm. This is in 
strong contrast to almost all other metals. Typically, a metal 
with a very large resistivity shows resistivity saturation\cite{Fisk}. 
Thus when $\rho$ reaches values of the order 0.1 m$\Omega$cm, the 
slope of $\rho(T)$ is typically reduced substantially, although not 
necessarily to zero. This happens when the apparent mean free path $l$
becomes comparable to the separation, $d$, of two atoms, the Ioffe-Regel 
condition\cite{ioffe}. This kind of behavior has been observed for many 
metals and it used to be considered a universal behavior\cite{Allen}.

The metals mentioned above and the alkali-doped 
fullerenes\cite{Hebard}
are apparent exceptions to this saturation behavior, with the
cuprates forming a particular challenge\cite{Allennv,Allenelph}.
The deviation from the ``universal'' behavior can be illustrated 
by considering Ioffe-Regel condition for La$_{2-x}$Sr$_x$CuO$_4$.    
We assume a ``large'' Fermi surface of cylindrical shape, 
containing $1-x$ electrons (or $1+x$ holes)\cite{fermisurface}. 
Assuming that $l=a$, where $a$ is the lattice parameter in the 
CuO$_2$ plane, we obtain the saturation resistivity
\begin{equation}\label{eq:0}
\rho_{\rm sat}={0.7\over \sqrt{1\pm x}} {\rm m}\Omega{\rm cm}.
\end{equation}
If we instead assume a ``small'' Fermi surface, $\sqrt{1\pm x}$
in the denominator is replaced by $\sqrt{x}$. In either
case, the experimental resistivity for small $x$ is much larger
than the saturation resistivity above, supporting the conclusion
that there is no saturation in these systems.

Here we demonstrate that the experimental data for the strongly
correlated high $T_c$ cuprates are, nevertheless, consistent 
with saturation. The saturation resistivity
can, however, be much larger than the Ioffe-Regel value.     
This is not entirely surprising, since the Ioffe-Regel condition
is based on a semiclassical picture, which is not valid when
$l\sim d$. The good agreement between the Ioffe-Regel condition
and the saturation for weakly correlated transition metal
compounds is therefore somewhat accidental. The large saturation
resistivity is shown to be due to a large reduction of the kinetic 
energy, due to strong correlation effects. Since the alkali-doped
fullerenes appear to lack saturation\cite{OG}, this puts these 
compounds in a special class, different from the cuprates.

To discuss resistivity saturation, we use the f-sum rule in a 
form appropriate for the models discussed here (only nearest neighbor 
hopping and no on-site matrix elements of the current 
operator)\cite{Maldague,advphys}
\begin{equation}\label{eq:1}
{2\over \pi}\int_0^{\infty} \sigma(\omega)d \omega=
-{1\over 2}{d^2e^2 \over N\Omega \hbar^2} \langle T_K\rangle,        
\end{equation}
where $\sigma(\omega)$ is the optical conductivity, $d$ is the
separation of the sites, $\Omega$
is the volume of a unit cell, $N$ is the number of unit cells and
$\langle T_K \rangle$ is the expectation value of the kinetic energy
operator. The prefactor $1/2$ refers to the two-dimensional case
and is replaced by $1/3$ for the three-dimensional case. To obtain 
an approximate upper limit to the resistivity, we assume that the 
(Drude) peak at $\omega=0$ has been smeared out and that $\sigma
(\omega)$ is a smooth function. The removal of the $\omega=0$ peak 
may be due to any scattering mechanism, electron-phonon\cite{Allenelph}, 
electron-electron or impurity scattering and, depending on the system, 
it may or may not happen at large values of $T$. We emphasize that 
our theory below does not depend on any particular scattering mechanism. 
We furthermore assume that $\sigma(\omega)=0$ for $|\omega|>W$, where 
$W$ is the one-particle band width. This should be a good approximation, 
since $|\omega|>W$ would involve multiple electron-hole pair excitations
and have a small weight. Explicit calculations for a model of transition
metal compounds and for the $t-J$ model confirm this. 
If $\sigma(\omega)$ had a box shape, $\sigma(\omega)\equiv 
\sigma(0)$ for $0\le |\omega| \le W$, the integral over $\sigma(\omega)$
would be $\sigma(0)W$ and we could write
\begin{equation}\label{eq:2}
\sigma(\omega=0)={\gamma \hbar \over W}\int_0^{\infty} 
\sigma(\omega)d \omega,   
\end{equation}
with $\gamma=1$. For a more realistic shape of $\sigma(\omega)$,
where $\sigma(\omega)$ has a maximum for $\omega=0$, it follows that 
$\gamma>1$. Following the considerations for a model of transition metal
compounds with electron-phonon scattering\cite{TM}, we may set 
$\gamma=1.9$.  In the $t-J$ model, considering electron-electron 
scattering, we find that $\sigma(\omega)$ is more concentrated 
to small $\omega$, giving a larger value of $\gamma$.
If the electrons would tend to localize, $\gamma$ could be much 
smaller than unity. This should, however, not happen at the large values 
of $T$ considered here. Measurements\cite{Takenaka} of $\sigma(\omega)$ 
for La$_{2-x}$Sr$_x$CuO$_4$ at large $T$ show a similar behavior as 
assumed in Eq. (\ref{eq:2}), apart from some sharp phonon structures 
at finite $\omega$.    

We first consider noninteracting electrons in a band with the
orbital degeneracy $n$. The scattering is assumed to be due to 
the electron-phonon interaction. For $T\ll W$ and neglecting the 
influence of the electron-phonon coupling on the kinetic energy,
\begin{equation}\label{eq:3}
\langle T_K\rangle =2n\int_{-W/2}^{\mu}\varepsilon
N(\varepsilon)d\varepsilon\equiv -2n\alpha W N,
\end{equation}
where $N(\varepsilon)$ is the density of states per orbital 
and spin, $\mu$ is the 
chemical potential and $\alpha$ depends on the shape $N(\varepsilon)$ 
and the filling. For semi-elliptical $N(\varepsilon)$ and filling 0.4,  
$\alpha=0.10$. This result is relatively independent of the specific
shape of $N(\varepsilon)$, and using, e.g., a constant $N(\varepsilon)$
leads to $\alpha=0.12$.  For a three-dimensional system we then obtain
\begin{equation}\label{eq:4} 
\sigma(0)={\pi n\over 3}\alpha\gamma 
{d^3\over \Omega}{e^2\over \hbar d},
\end{equation}
which provides an approximate upper limit to the resistivity for 
$T\ll W$. Saturation happens if $\rho(T)$ grows so rapidly that
this approximate upper limit is approached for experimentally 
accessible values of $T$. This happens for some transition metal 
compounds, with the A15 compounds, such as Nb$_3$Sb, being particularly
striking examples. The important orbitals in Nb$_3$Sb are the Nb 
$d$-orbitals, leading to $n=5$. Considering the A15 lattice of 
Nb$_3$Sb, we obtain the saturation resistivity 0.14 m$\Omega$cm, 
which compares well with the experimental saturation resistivity of
about 0.15 m$\Omega$cm\cite{Fisk}. This illustrates the
usefulness of this approach for obtaining a saturation 
resistivity.

We next consider the strongly correlated cuprates. To 
calculate the kinetic energy $\langle T_K\rangle$, we need     
an appropriate model for these systems. The transport
properties should mainly be determined by the Cu-O antibonding
band of Cu $x^2-y^2$ and O $2p$ character. Thus we consider
one orbital per site and the orbital degeneracy $n=1$. These orbitals
are put on a two-dimensional square lattice, and we consider a 
a layered system with the lattice parameters ($a$,$a$,$2c$).  We 
further assume that the Coulomb interaction $U$ between two 
electrons on the same site is so large that double occupancy of a 
site can be neglected. This can be described by the $t-J$ 
model\cite{Rice}.  The properties of this model have been studied 
extensively by Jaklic and Prelovsek\cite{advphys}. The system has 
the hole doping $x$ $(>0)$. We emphasize that the $t-J$ model is 
only used to estimate $\langle T_K\rangle$. We do not exclude 
other scattering mechanisms; we just neglect their contribution 
to the kinetic energy.

For simplicity, we first consider $J=0$.
By expanding in $1/T$, we obtain the large $T$ limit of the kinetic 
energy as         
\begin{equation}\label{eq:7}
\langle T_K\rangle ={{\rm Tr} \ T_Ke^{-H/T}\over {\rm Tr}\ e^{-H/T}}
\approx -{1\over T} \lbrack {{\rm Tr} \ T_K^2\over {\rm Tr} \ 1}
\rbrack_{\rm no \ doubble \ occ},
\end{equation}
where the right hand site is evaluated for states without double 
occupancy. Since the expression contains $T_K^2$, this involves 
a hole hopping to a neighboring site and back. In the $t-J$ model
this is only possible if the neighboring site has no holes. The 
probability for having a hole or no hole on a given site is $x$ and 
$(1-x)$, respectively. The probability for hopping between two nearest
neighbor sites is then approximately $x(1-x)$. Summing over the four 
nearest neighbors, this gives the large $T$ limit
\begin{equation}\label{eq:8}
\langle T_K\rangle =-4Nt^2{x(1-x)\over T}.                          
\end{equation}

\begin{figure}[bt]
{\rotatebox{-90}{\resizebox{!}{3.5in}{\includegraphics
{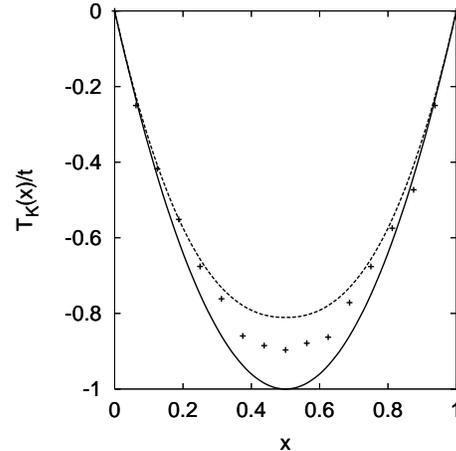}}}}
\caption[]{\label{fig:0} The kinetic energy $T_K(x)$ as a function of 
doping $x$ in the $t-J$ model for $J=0$. The crosses show the result of exact
diagonalization for $4\times 4$ two-dimensional cluster\cite{tmodel},
the full curve Eq. (\ref{eq:9}) and the broken curve the result for 
spinless fermions\cite{Emery,Calandra}.  The figure illustrates the 
approximate $x(1-x)$ behavior of the kinetic energy.  }
\end{figure}

Using similar arguments, we can obtain a rough estimate for 
the $T=0$ kinetic energy
\begin{equation}\label{eq:9}
\lbrack \langle T_K \rangle\rbrack_{T=0}=-4tx(1-x)N.
\end{equation}
This result is shown by the full curve in Fig. \ref{fig:0}.
A similar result, but with a 
smaller coefficient, can be obtained variationally, by assuming 
that the holes behave as spinless fermions in a ferromagnetic 
background (broken curve in Fig. \ref{fig:0})\cite{Emery,Calandra}. 
These results can be compared with results from an exact 
diagonalization\cite{tmodel} of a $4\times 4$ $t-J$ model with 
$J=0$ (crosses in Fig. \ref{fig:0}). The figure illustrates the 
approximate $x(1-x)$ dependence of the kinetic energy. 
In the following we use the $t-J$ model with $J/t=0.3$ and 
$t=0.4$ eV\cite{advphys}.  We solve the model,  using the finite $T$ exact 
diagonalization technique of Jaklic and Prelovsek\cite{Prelovsek}. 
We consider a finite cluster of $4\times 4$ atoms and use the grand
canonical ensemble. The resulting numerical results for the kinetic 
energy for small values of $T$ and $x$ can be approximately expressed in the 
form of Eq. (\ref{eq:9}), but with the prefactor 4 reduced to about 3.4.
For the values of $T$ of interest here, the $T$ dependence of 
$\langle T_K \rangle$ is weak, and it is neglected in the following. 
If the system has a strong electron-phonon interaction, the kinetic 
energy should be somewhat further reduced.

Combining Eqs. (\ref{eq:1}, \ref{eq:9}), we predict that the 
right hand-side of Eq. (\ref{eq:1}) should be proportional to $x$,
for small $x$. The effective number of carriers per $\rm Cu$ atom 
$N_{eff}(x)$ can be estimated integrating the optical conductivity up to 
$\omega_c=1.2$ eV, as suggested by Yamada {\it et al.}\cite{transition}.
This also includes some contribution from interband transitions, 
and Yamada {\it et al.}\cite{transition} therefore subtracted
$N_{eff}(0)$, which should contain only interband transitions.
They found that $N_{eff}(x)-N_{eff}(0)$ is indeed approximately 
proportional to $x$, up to $x=0.12$. Based on the absolute numbers 
by Uchida {\it et al.}\cite{Uchida}, we find that our factor of 
proportionality in Eqs. (\ref{eq:1}, \ref{eq:9}) agrees with experiment 
to within about 20-30 $\%$. Due to the problems of removing      
the interband transitions, this should be a good agreement.

We consider conduction in the ab-plane of a cuprate and assume 
that the $\omega=0$ peak is gone. Insertion of Eq. (\ref{eq:9}), 
but with the prefactor 3.4, in Eqs. (\ref{eq:1}, \ref{eq:2}) gives 
an approximate upper limit to the resistivity for small and intermediate
values of $T$
\begin{equation}\label{eq:10}
\rho={0.4\over x(1-x)} \hskip0.3cm {\rm m}\Omega{\rm cm},
\end{equation}
where we have used distance $c=6.4$ \AA \ between the CuO$_2$ planes.
Inserting the calculated kinetic energy for the $t-J$ model would 
lead to a somewhat larger resistivity for small $x$ and it would 
introduce a weak $T$ dependence. The saturation resistivity in Eq. 
(\ref{eq:10}) is  much larger than Eq. (\ref{eq:0}) for small $x$.
It illustrates that one should expect saturation at a substantially 
larger resistivity than predicted by the Ioffe-Regel condition.
The saturation resistivity in Eq. (\ref{eq:10}) is also much larger 
than the result derived\cite{TM} for a model of weakly correlated 
transition metal compounds
\begin{equation}\label{eq:11}
\rho \sim {0.2 d\over n}\approx {0.5 \over n},
\end{equation}
where $d$ is expressed in \AA.
This is partly due to the degeneracy being just $n=1$ for the $t-J$ 
model but $n=5$ for our model of the weakly correlated transition
metal compounds. Furthermore, the 
strong correlation drastically reduces the kinetic energy in the 
cuprates, which shows up as the factor $x(1-x)$ in the denominator 
of Eq. (\ref{eq:10}). Finally, the large value of the lattice 
parameter $c$ also increases $\rho$ for the cuprates. 

\begin{figure}[bt]
{\rotatebox{-90}{\resizebox{!}{3.5in}{\includegraphics
{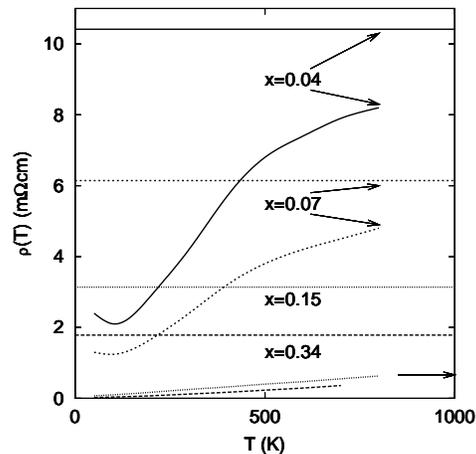}}}}
\caption[]{\label{fig:1} The resistivity as a function of $T$ for
La$_{2-x}$Sr$_x$CuO$_4$\cite{hightcres} and the saturation 
resistivity according Eq. (\ref{eq:10}) for $x=0.04$ (full curve) $x=0.07$
(broken curve) $x=0.15$ (dotted curve) and $x=0.34$ (chain curve). 
The horizontal arrow shows the saturation resistivity expected 
from the Ioffe-Regel condition.  The figure illustrates that
there are signs of saturation for small $x$ at roughly those 
resistivities where saturation is expected. For larger $x$, much 
larger $T$ would have to be considered to test whether there 
is saturation.}
\end{figure}

We now consider the $T$ dependent resistivity for different
doping. The experimental results of Takagi {\it et al.}\cite{hightcres}
are shown in Fig. \ref{fig:1}. For $T\le 300$ K, more recent 
results have been obtained by Ando {\it et al.}\cite{Ando}. 
Their resistivities are qualitatively similar but generally 
smaller than the results in Fig. \ref{fig:1}.  The results by 
Takagi {\it et al.}\cite{hightcres} are compared with the saturation 
resistivity in Eq. (\ref{eq:10}). For all values of $x$, the 
experimental resistivity is smaller
than the saturation resistivity predicted here. The experimental
results do therefore not demonstrate absence of saturation. Actually,
for $x=0.04$ and $x=0.07$ the data show signs of saturation for 
values not much smaller than the expected saturation resistivities, 
but at much larger values than the saturation resistivity  predicted 
by the Ioffe-Regel criterion (Eq. (\ref{eq:0})) (indicated by an arrow). 
This shows that the Ioffe-Regel criterion is invalid for these systems.
It is interesting to plot $x\rho(T)$\cite{Batlogg}. In such a plot the 
saturation resistivity becomes almost independent of $x$. Indeed the 
curves of $x\rho(T)$ for $x=0.04$ and $x=0.07$ fall almost on top of 
each other, suggesting that close to saturation there is a scaling of 
$\rho(T)$ by $1/x$, as implied by the arguments above. 
For $x=0.15$ and $x=0.34$ one would have to study much higher (and 
experimentally unaccessible) values of $T$ to determine whether or 
not there is saturation.  

La$_{2-x}$Sr$_x$CuO$_4$ has a structural transition from an 
orthorhombic to a tetragonal phase. This transition happens,
however, at a somewhat lower temperature\cite{transition}
than the break in the slopes of the curves
for $x=0.04$ and $x=0.07$ in Fig. \ref{fig:1}. 

We have also considered the resistivity of
YBa$_2$Cu$_3$O$_{6+\delta}$\cite{ybacuo}, 
Nd$_{2-x}$Ce$_x$CuO$_{4-y}$\cite{ndcecuo}, 
Bi$_2$Sr$_2$Ca$_{1-x}$Y$_x$Cu$_2$O$_{8+y}$\cite{bisrcaycuo,steglich},
Tl$_2$Ba$_2$CuO$_{6+\delta}$\cite{tlbacuo}
and HgBa$_2$Ca$_{n-1}$Cu$_n$O$_{2n+2+x}$\cite{hgbacacuo}. In all these
cases we find that the experimental resistivity does not
exceed the expected saturation resistivity. In the case of 
Bi$_2$Sr$_2$Ca$_{1-x}$Y$_x$Cu$_2$O$_{8+y}$ saturation has been
reported\cite{steglich}. This provides further
support for our theory.

Above we have provided an upper limit to the resistivity
based on the f-sum rule and the calculated kinetic energy
and without specifying a scattering mechanism.
We have also calculated the resistivity of the $t-J$ model directly,
although the results are not very accurate for the small clusters 
considered here.  For $x=0.15$, the calculated resistivity is of the 
right order of magnitude, while for $x=0.07$ and $x=0.04$ it is  
substantially smaller than the experimental results in Fig. 2. 
This suggests that there is an additional scattering mechanism, beyond
the electron-electron scattering included explicitly in the $t-J$ model, 
which becomes important for $x=0.04$ and $x=0.07$. This additional 
mechanism apparently makes the resistivity increase so rapidly for 
small $T$ that the approximate upper limit (\ref{eq:10}) is approached 
and the resistivity shows sign of saturation. The estimate of the 
upper limit (\ref{eq:10}) is nevertheless correct, unless the additional 
scattering mechanism influences the kinetic energy in a substantial way. 

Above we have discussed the situation when $T$ is much smaller than 
the band width, which applies for values of $T$ which can be reached
experimentally. The kinetic energy $T_K$ then has a weak $T$ 
dependence, leading to an upper limit for the resistivity which
is only weakly $T$ dependent. As can be seen from Eq. (\ref{eq:8}), however, 
the $T$ dependence becomes strong for very large $T$ ($>W/4$). 
Inserting Eq. (\ref{eq:8}) into Eqs. (\ref{eq:1},\ref{eq:2}) then 
leads to an upper limit for $\rho(T)$ which grows linearly with $T$. 
A similar result would be found even for noninteracting electrons 
scattered by impurities, i.e., for a $T$-independent mechanism,
where the $T$ dependence can then be traced back to the $T$ dependence
of the Fermi functions. This illustrates that ``saturation'' does
not necessarily imply that the resistivity becomes $T$ independent,
just that its slope is reduced substantially at large $T$\cite{Allen}. 
The effects discussed in this paragraph, however, refer to very 
large values of $T$, of the order of 10000 K. They are therefore 
only of academic interest, unless the band width is very small.

To summarize, we have shown that the experimental resistivity
data for a number of cuprates are consistent with resistivity
saturation. The saturation resistivity is, however, much larger
than for, e.g., A15 compounds, due to the strong suppression
of the kinetic energy in the strongly correlated cuprates.
The Ioffe-Regel condition is therefore  violated for a
many of these systems. 

We would like to thank P. Horsch and B. Keimer for useful discussions, 
and the Max-Planck-Forschungspreis for financial support.

\end{multicols}

\end{document}